\documentstyle[12pt,fleqn,epsfig]{article}
\newcommand{\be}{\begin{equation}}
\newcommand{\ee}{\end{equation}}
\newcommand{\bea}{\begin{eqnarray}}
\newcommand{\eea}{\end{eqnarray}}

\begin{document}

\date{}
\title{Medium effects in the production and $\pi^0 \gamma$ decay of
$\omega$-mesons in $pA$ collisions in the GeV region
 \footnote{Supported by DFG and RFFI}}
\author{Ye.S. Golubeva$^1$, L.A. Kondratyuk$^2$, M. B\"uscher$^3$, \\
W. Cassing$^4$, V. Hejny$^3$ and H. Str\"oher$^3$} \maketitle

\begin{center}
$^1$ Institute of Nuclear Research, 117312 Moscow, Russia \\

$^2$ Institute of Theoretical and Experimental Physics, 117259
Moscow, Russia \\

$^3$ Forschungszentrum J\"ulich, Institut f\"ur Kernphysik, 52425
J\"ulich,
 Germany \\

$^4$ Institute for Theoretical Physics, University of Giessen,
D-35392 Giessen, Germany \\
\end{center}

\begin{abstract}
The $\omega$-resonance production and its $\pi^0 \gamma$ decay in
$p- A$ reactions close to threshold is considered within the
Intranuclear Cascade  (INC) Model. The $\pi^0 \gamma$ invariant
mass distribution shows two components which correspond to the
$\omega$ decay 'inside' and 'outside' the nucleus, respectively.
The 'inside' component is distorted by medium effects, which
introduce a mass shift as well as collisional broadening for the
$\omega$-meson and its decaying pion. The relative contribution of
the 'inside' component is analyzed in detail for different
kinematical conditions and nuclear targets. It is demonstrated
that a measurement of the correlation in azimuthal angle between
the $\pi^0$ and $\gamma$ momenta allows to separate events related
to the 'inside' $\omega$ decay from different sources of
background when uncorrelated $\pi^0$'s and $\gamma$'s are
produced.
\end{abstract}


\vspace{2cm} \noindent PACS: 25.40.Ep; 25.80.-e, 25.80.Ek


\newpage

\section{Introduction}


The question about the modifications of the vector meson
properties in the nuclear medium has received a vivid attention
during the last years (cf. Refs.
\cite{1,2,Ca95,Klingl,3,4,5,6,7,8,9,Golub}). Whereas the
$\rho$-meson is expected to practically dissolve already at normal
nuclear matter density $\rho_0$, the $\omega$-meson is expected to
survive as a quasi-particle at densities $\leq \rho_0$, i.e. in
ordinary nuclei. A meson or baryon resonance produced in a pion or
proton induced reaction on nuclei can decay 'outside' or 'inside'
the target nucleus. Correspondingly, the invariant mass
distribution of the decay products for each resonance contains two
components \cite{6,Golub}. The first component is described by a
Breit-Wigner formula with vacuum properties of the resonance mass
$M_0$ and width $\Gamma _0$. The second component is
strongly distorted by the nuclear medium due to a collisional broadening $%
\delta \Gamma $ and a possible mass shift $\delta M$ , where in
first approximation $\delta \Gamma $ and $\delta M$ are
proportional to the nuclear density. In principle, medium
modifications of $\omega $ mesons can be detected directly by
measuring the dilepton invariant mass spectra in $hA$ and $AA$
collisions (see e.g. Refs. \cite{Golub,Schoen} and references
therein). The advantages of this method are related to the fact
that the dilepton mass spectra are almost undistorted by the final
state interactions. However, the $\omega $-signal in the dilepton
mode is rather weak ($BR(\omega \to e^+ e^-)\simeq 7.1\times
10^{-5}$) and is always accompanied by a comparatively large
background from $\rho ^0$ decays. From this point of view it is
useful  to consider also the $\omega \rightarrow \pi ^0\gamma $
decay which has a branching ratio of about 3 orders of magnitude
higher (see Refs. \cite{NAN-95,Sibirtsev}). As it was shown in
Ref. \cite{Sibirtsev} the effect of the $\omega $ mass shift in
the $\pi ^0\gamma $ invariant mass distribution can be identified
on top of the background which is related to $\pi ^0 N$
rescattering events.

In the present paper we discuss in more detail the in-medium
$\omega$ decay to an off-shell $\pi^0$ and a photon and
investigate additional criteria to separate the $\omega
\rightarrow \pi ^0\gamma $ signal in $pA$ collisions at GeV
energies from different sources of background such as $\pi^0
\pi^0$ or $\pi^0 \eta$ when one photon from the meson decays is
not detected or misidentified..

The paper is organized as follows: In Section 2 we will briefly
discuss the propagation of a resonance in the nuclear medium, its
decay to an off-shell pion and a photon and describe the
Intranuclear Cascade (INC) Model that is used in the simulation of
events for the reaction $pA \to \omega (\pi^0 \gamma) X$. In
Section 3 we present the results of our calculations and summarize
our work in Section 4.

\section{Theoretical framework}

\subsection{Propagation of a resonance in the nuclear medium}

\indent

We here consider the production of $\omega$-resonances in $pA$-
collisions not far above the threshold. Due to the kinematics of
the production process the resonance will be quite fast in the
laboratory frame (with the target almost at rest) and its
propagation through the nucleus can be described within the
framework of the eikonal approximation. The corresponding Green
function, which describes the propagation of the resonance from
the point $\vec{r}=(\vec{b},z)$ (where it is produced) to the
point $\vec{r}^{\prime}=(\vec{b}^{\prime},z^{\prime})$ (where it
decays or rescatters incoherently) can be written as
\cite{6,Golub}
\begin{equation}  \label{Green}
G_p(\vec{b}^{\prime},z^{\prime};\vec{b},z)=\frac{1}{2ip} \exp
\{i\int^{z^{\prime}}_z[p+{\frac{1}{2p}(\Delta +4\pi f(0)\rho_A(\vec{b}}%
,\zeta))]d\zeta\} \times
\delta(\vec{b}-\vec{b}^{\prime})\theta(z^{\prime}-z),
\end{equation}
where the $z$-axis is directed along the resonance momentum $\vec{p}$,
$\vec{b}$ is the impact parameter and
\begin{equation}  \label{delta}
\Delta = P^2- M^2_R - iM_R \Gamma_R
\end{equation}
is the inverse resonance propagator, while $M_R$ and $\Gamma_R$
are the resonance mass and width in the vacuum. The four-momentum
$P$ in (\ref{delta}) can be defined through the four-momenta of
the resonance decay products,
\begin{equation}
P=p^*(\pi^0) + p(\gamma ),
\end{equation}
where $p^*(\pi^0)$ is the pion 4-momentum in the medium (cf.
Section 2.2). Eq. (\ref{Green}) is written in the low density
approximation with the optical potential defined as (see, e.g.
Ref. \cite{ericson})
\begin{equation}  \label{optpot}
U(\vec{r})= - 4\pi f(0)\rho_A(\vec{r}) ,
\end{equation}
where $f(0)$ is the forward $\omega N$-scattering amplitude and
$\rho_A(\vec{r})$ is the nuclear density.

The corresponding amplitude, which describes the production of the
resonance in the point $\vec{r}=(\vec{b}%
,z) $ inside the nucleus and its decay in the point $\vec{%
r}^{\prime}=(\vec{b}^{\prime},z^{\prime})$, can be written as
\begin{equation}  \label{both}
M(\vec{P},P^2;\vec{b},z)=B f_{pN_j \to \omega X_i } \int d^2\vec{b^{\prime}}
dz^{\prime}\{\exp(-i \vec{p} \vec{r} ^{\prime}) G_p(\vec{r} ^{\prime}-\vec{r}%
) \ \exp(i \vec{p} \vec{r})\} f_{\omega \to \pi^0 \gamma}.
\end{equation}
Here $f_{pN_j \to \omega X_i}$ is the production amplitude; $X_i
=N N, \pi NN, \Delta N,...$; $f_{\omega \to \pi^0 \gamma}$ is the
decay amplitude and $B$ is a normalization factor which takes into
account the attenuation of the initial proton flux due to the
screening from other nucleons.

Since we consider energies not far above threshold for $\omega$
production, the $\omega$ can be produced in the first hard
proton-nucleon collision or through the two-step mechanism with an
intermediate production of a pion \cite{14,INC2}: $pN_j\to \pi pN,
\pi N_k \to \omega N$. The contribution of the two-step scattering
mechanism has to be added to the direct mechanism described by Eq.
(5). In the latter case the production amplitude is obtained by
replacing $f_{pN_j \to \omega X_i }$ in (5) by $f_{\pi N_k \to
\omega X_i }$ and $G_p$ by the pion propagator $G_\pi$. In
addition an integration over all pion production points has to be
performed.

If the resonance decays outside the nucleus, the amplitude for
$\pi^0 \gamma$ production -- in case of the primary production
channel -- can be written as
\begin{equation}  \label{both1}
M^i_j(\vec{P}, P^2;\vec{b},z) = B  f(pN_j \to \omega X_i)\{A_{in}(\vec{P}%
,P^2; \vec{b},z)+ A_{out}(\vec{P}, P^2; \vec{b},z)\} f_{\omega \to
\pi^0\gamma},
\end{equation}
where the contributions from the resonance decays 'inside' and
'outside' the nucleus can be expressed as
\begin{equation}  \label{in}
A_{in}(\vec{P}, P^2; \vec{b}, z)=\frac{1-\exp[i(\Delta^*/2k)(z_s-z)]}{\Delta^*%
},
\end{equation}
and
\begin{equation}
A_{out}=\frac{\exp[i(\Delta^*/2k)(z_s-z)]}{\Delta}
\end{equation}
with
 \begin{equation}
 \label{star}
\Delta^*=\Delta+4\pi f(0)\rho_0= P^2-M^{*2}_R+iM_R^*\Gamma_R^* .
\end{equation}
Eq. (\ref{star}) describes  the resonance propagator
$(\Delta^*)^{-1}$  in the nuclear medium with distorted values of
the resonance mass ($M_R*$) and width ($\Gamma_R^*$) defined by
\begin{equation}
M_R^{*2}=M^2_R-4\pi Re{f(0)}\rho_A,
\end{equation}
\begin{equation}
M_R^*\Gamma^*_R=M_R\Gamma_R+4\pi Im{f(0)}\rho_A.
\end{equation}
When the resonance decays 'inside' the nucleus the amplitude
(\ref{both1}) contains only the 'inside' component $A_{in}$
defined by (\ref{in}). Evidently, the $\pi^0 \gamma$ invariant
mass spectrum is given by the following expression:
\begin{equation}  \label{sig}
\frac{d\sigma}{dM} = \sum_{i,j} \int d^2b dz \ \rho(\vec{b},z) \ |M^i_j(\vec{%
P},\vec{P}^2,\vec{b},z)|^2 ,
\end{equation}
where the sum is taken over all nucleons in the target and all production
channels, respectively.

\subsection{$\omega$ decay to off-shell pions}
Since not only the $\omega$-meson, but also the decay pion changes
its spectral function in the medium, the in-medium $\omega$ Dalitz
decay has to be discussed explicitly. Now let the in-medium
$\omega$ mass be $M$ and the decay pion have a mass $m^*_\pi$ that
might be selected by Monte Carlo according to its in-medium
spectral function with width $\Gamma_\pi^*$ and mass shift $\delta
m_\pi$. Energy and momentum conservation - in the rest frame of
the $\omega$ meson - than implies
\begin{equation}
M^2 = (E_\pi^*+E_\gamma)^2 = (\sqrt{p_{\pi}^{*2} + m_{\pi}^{*2}} +
E_\gamma)^2  \label{e1},
\end{equation}
with $E_\gamma = |p_\pi^*|$ denoting the photon energy which in
magnitude equals the momentum of the decay pion. Whereas the
photon propagates to the vacuum without distortion, the in-medium
pion changes its momentum and spectral function during the
propagation to the vacuum according to quantum off-shell
propagation \cite{ca1,ca2,ca3}. In the particular case, where the
pion self energy $\Sigma_\pi$ has no explicit time dependence
($\partial_t \Sigma_\pi = 0$) and is only a function of momentum
and density, which holds well for the case of $p + A$ reactions,
the energy of the pion $E_\pi^* = E_\pi$ is a constant of motion
(cf. Eq. (20) in Ref. \cite{ca2}). Thus the off-shell mass and
momentum balance out during the propagation as shown graphically
in Ref. \cite{ca2} in Figs. 1 and 2 for a related problem. The
magnitude of the pion momentum in vacuum - if not scattered
explicitly or being absorped - then is simply given by
\begin{equation}
p_\pi^v = \sqrt{p_\pi^{*2} + m_\pi^{*2} - m_0^2}, \label{e2}
\end{equation}
where $m_0$ denotes the vacuum pion mass and $p_\pi^*$ the pion
momentum in the $\omega$ decay in-medium. The invariant mass
(squared) of the pion and photon in the vacuum then is given by
\begin{equation}
M_r^2 = (E_\pi + E_\gamma)^2 - (p_\pi^v - E_\gamma)^2 = M^2 -
(p_\pi^v - E_\gamma)^2 . \label{e3}
\end{equation}
The pion propagation thus leads to a slight downward shift of the
invariant mass in vacuum relative to its original in-medium value.
We note, that the above considerations apply well for heavy nuclei
where the corrections in energy due to the recoil momentum of the
nucleus ($|p_\pi^v - E_\gamma|$) can be discarded.

For an actual quantification of this effect one has to specify the
pion spectral function e.g. at nuclear matter density (assuming a
linear dependence on density $\rho$ for simplicity). The pions,
that originate from the $\omega$ decay in $p+A$ reactions, have
typical momenta $\geq 300$ MeV such that their interaction cross
section with nucleons is in the order of 30-35 mb, while their
velocity $\beta_\pi$ relative to the target is close to 1.
Consequently the collisional broadening can be estimated as
\begin{equation}
\delta \Gamma_\pi \approx \beta_\pi \sigma _{\pi N} \rho_0 \approx
100 - 120 \ MeV, \label{e4} \end{equation}
 which is still quite substantial. For the
$\omega$ in-medium spectral function we assume a Breight-Wigner of
width $\Gamma_{tot}$ = 50 MeV and pole mass of 0.65 GeV (cf. solid
line in Fig. 1). Now for each invariant mass $M$ the decay of the
$\omega$ to a photon and an in-medium pion can be evaluated and
the pion momentum $p_\pi^V$ according to (\ref{e2}). To obtain
some upper limit for the pion off-shell effects we take $\delta
\Gamma_\pi$ = 200 MeV.  The reconstructed invariant mass
distribution in (\ref{e3}) is shown in Fig. 1 by the dashed line,
which is close to the original $\omega$ spectral function (solid
line) and indicates that such corrections can be safely neglected
in view of presently achievable experimental mass resolutions.
Furthermore, the effects are most pronounced for very low
invariant masses $M$ where, however, the spectrum is strongly
distorted by $\pi^0$ rescattering (see below). Thus we continue
our calculations with on-shell spectral functions for the pions
without substantial loss in accuracy. For the general case of
off-shell dynamics in transport or cascade calculations we refer
the reader to Refs. \cite{ca1,ca2,ca3}.

\subsection{The Intranuclear Cascade model and interaction parameters}


The yields of $\omega$-mesons are calculated within the framework
of the intranuclear cascade model (INC) developed in Ref. \cite
{INC1,INC2}, which was also used in Ref. \cite{Golub} for the
analysis of medium effects for $\omega$'s produced in pion-nucleus
reactions with respect to their dilepton decay. In the INC the
linearized kinetic equation for the many-body distribution
function  is solved numerically by assuming that during the
evolution of the cascade the properties of the target nucleus
remain unchanged. Within the INC approach the target nucleus is
regarded as a mixture of degenerate neutron and proton Fermi gases
in a spherical potential well with a diffuse surface. The momentum
distribution of the nucleons is treated in the local density
approximation for a Fermi gas.

The following inputs were used for the elementary cross sections
(see also Refs. \cite{Golub,INC2}: i) For the cross section of the
reaction $pN \to \omega p N$ we use parametrization from Ref.
\cite{14};\\ ii) For the process $\pi^- p \to \omega n $ we use a
parametrization of the experimental data from Refs.
\cite{INC2,Cugnon}:
\begin{equation}
\sigma(\pi^- p \to \omega n)= C \frac{P_{\pi N}-P_{\omega}^0}{P_{\pi
N}^{\alpha}-d},
\end{equation}
where $P_{\pi N}$ is the relative momentum (in GeV) of the pion-nucleon pair
while $P_{\omega}^0$ = 1.095 GeV is the threshold value. The parameters $C =
$ 13.76 mb$\cdot $GeV/c$^{\alpha -1}$, $\alpha$=3.33 and $d$=1.07 GeV$^{\alpha}$
describe satisfactorily the data on the energy-dependent cross section
 $\pi^- p \to \omega n $  in the near-threshold energy region.\\
iii) For the $\omega N$ elastic and total cross sections we use
the same parametrization as in Ref. \cite{Golub,INC2}.
 The angular distributions for the particles produced in the elementary
reactions are considered to be isotropic in the corresponding
c.m.s. since we operate close to threshold energies.

The propagation of vector mesons in the nuclear medium is
described in the same manner as in our previous works
\cite{Golub}. A collisional broadening of the resonance in the
nuclear medium is introduced according to Eq. (11) where $Im f
(0)$ has been expressed through $\sigma_{tot}(\omega N)$ using the
optical theorem. The $\omega$ mass shift is modeled according to
Hatsuda and Lee \cite{3} as
\begin{equation}  \label{Brown}
M^*_R=M_R + \delta M_R=M_R(1 - \alpha \frac{\rho_A(r)}{\rho_0}),
\end{equation}
where $\rho_A (r)$ is the nuclear density at the resonance decay point,
$\rho_0 = 0.16 fm^{-3}$, and $\alpha = 0.18$.

The decay of the $\omega$ to $\pi^0 \gamma$ with its actual spectral shape is
performed as in Refs. \cite{Golub} : when the resonance decays
inside the nucleus its mass is generated according to a Breit--Wigner
distribution with average mass $M^*_R$ (\ref{Brown}) and $%
\Gamma^*_R=\Gamma_R+\delta \Gamma_R$ (see (11)), where the collisional
broadening and the mass shift are calculated according to the local nuclear
density. Its decay to $\pi^0 \gamma$ is recorded as a function of the
corresponding invariant mass bin and the local nucleon density $\rho_A$. If
the resonance leaves the nucleus, its spectral function automatically
coincides with the free distribution because $\delta M_R$ and $\delta
\Gamma_R$ are zero in this case. Secondary interactions of $\pi^0$
from $\omega$ decay are taken into account using realistic $\pi^0 N$ elastic,
inelastic and charge exchange cross sections.

\section{Production and $\pi^0 \gamma $ decay of $\omega$ mesons}
\subsection{General aspects}
We continue with the actual numerical results of our calculations
and first have a look at the $\omega$ momentum distributions in
the laboratory for $p+A$ collisions at $T_p$ = 2.4 GeV. The solid
histograms in Fig. \ref{fig:om1} show the calculated momentum
distributions of $ \omega $ mesons in the reactions $p\ +^{12}C$
(left figures) and $p+^{208}Pb$ (right figures). Hatched areas in
the upper part of the figure describe the contributions from the
two-step $\omega $ production mechanism with an intermediate pion,
which have a maximum at slightly lower momenta than the total
spectrum (solid histograms) for both targets due to the different
kinematics. The dotted and dashed histograms in the lower part of
Fig. \ref{fig:om1} correspond to $\omega $
decays 'outside' ( $\rho \leq $ 0.03 $\rho _0$) and 'inside'
( $\rho \geq $0.03 $%
\rho _0$) a nucleus, respectively. It is clearly seen that in case
of the $^{12}C$ target most of the $\omega$-mesons decay in the
vacuum (dotted histogram) while in case of the $Pb$ target both
components (integrated over momentum) become roughly comparable.
By performing cuts on the $\omega$ momentum $p_\omega$ one can
vary systematically the fraction of 'inside' (dashed histogram) to
'outside' decays (dotted histogram). Furthermore, the hatched
areas in the lower part of Fig. \ref{fig:om1} represent the events
from the 'inside' component with surviving neutral pions, i.e.
with $\pi^0$'s which propagate to the vacuum. Note, that this
contribution includes also pions that rescatter elastically. The
latter component (of surviving neutral pions) is dominant for the
'inside' decay in case of the $^{12}C$ target, but reduces to
roughly 60\% for the $Pb$ target. Nevertheless, there is still a
large fraction of $\pi^0$'s that escape without rescattering,
since for $\omega$-mesons with momenta around 0.5 GeV/c the
decaying pions also have large momenta with respect to the target
(at rest in the laboratory) such that their interaction cross
section with nucleons is only about 30-35 mb.

The upper and lower histograms in Fig.  \ref{fig:om2} show the
$\pi ^0\gamma $ invariant mass spectra from $p+Pb$ collisions at
2.4 GeV  for slow ( $p_\omega \leq $0.5 GeV/c) and fast (
$p_\omega \geq $ 0.5 GeV/c) $\omega $-mesons, respectively. The
right (left) figures demonstrate the distributions with (without)
medium effects for the $\omega$ mesons taken into account whereas
the hatched areas correspond to $\omega $ decays outside the
nucleus. As seen from a comparison of the left and right figures,
the events with invariant mass $M(\pi^0 \gamma) \leq$ 0.75 GeV
come up not only due to a collisional broadening and/or mass shift
of $\omega$-mesons but also due to $\pi^0$ rescattering.
Especially for 'slow' $\omega$'s this $\pi^0$ rescattering
contribution is more pronounced since in this case the cross
section for $\pi^0$ rescattering is quite large such that the
interesting signal from the in-medium $\omega$ decay is masked
substantially.

The question now arises to what extent one might possibly
disentangle in-medium effects due to collisional broadening and/or
$\omega$-meson mass shifts, respectively. To this aim we have
performed calculations for a $Pb$ target (Fig. \ref{fig:om3}) and
a $^{12}C$ target (Fig. \ref{fig:om4}) at 2.4 GeV for different
assumptions on the in-medium properties of the $\omega$-meson. In
both figures we present again the $\pi^0 \gamma $ invariant mass
distributions, however, without performing cuts on the $\omega$
momentum.  Figs. \ref{fig:om3}a) and \ref{fig:om4}a) show the
invariant mass distributions when $\omega$ medium effects are
discarded and all the distorsion of the $M(\pi^0 \gamma)$ spectrum
from $\omega$ decay is due to $\pi^0$ rescattering. Parts b) of
Figs. \ref{fig:om3} and \ref{fig:om4} demonstrate the $\pi^0
\gamma$ invariant mass spectrum when only the collisional
broadening $\delta \Gamma$ is taken into account while in parts c)
both in-medium effects, i.e. $\delta \Gamma$ and $\delta M$ are
taken into account.  The hatched areas in parts a)-c) show the
events with rescattered pions, the dotted histograms in a)-c)
describe the events with $\omega$'s decaying outside, while the
dashed histograms correspond to $\omega$'s decaying inside the
nucleus with $\pi^0$ rescattering switched off, such that the
$\omega$ spectral function at the decay becomes visible
explicitly.

For an actual comparison of the different scenarios the parts d)
in Figs. \ref{fig:om3} and \ref{fig:om4} show the total spectra
from a) - c) again: the dashed  histograms correspond to the
calculations without in-medium effects (a), the dotted histograms
the  spectra including  collisional broadening $%
\delta \Gamma \neq 0$ but without a mass shift $\delta M =0$ (b),
while the solid histograms show the result with collisional
broadening $\delta \Gamma \neq 0$ and the mass shift $\delta M
\neq 0 $ (c). In the region $M(\pi^0 \gamma)$ = 0.65-0.75 GeV
there is a substantial enhancement in the spectrum due to medium
effects (solid histograms) relative to the vacuum $\omega$ decays
(dashed histogram) even when including $\pi^0$ rescattering. This
finding is in agreement with the results of previous calculations
on the $\omega$ properties from pion and proton induced reactions
on nuclei, that have investigated the $\omega$ dilepton or $\pi^0
\gamma$ decay \cite{Golub,NAN-95, Sibirtsev}, respectively.

\subsection{Background reduction}
Experiments on $\omega$-production in nuclei might be carried out at
the proton accelerator COSY-J\"ulich using a photon detector to look
for the 3$\gamma$ invariant mass distribution and selecting
2$\gamma$'s in the invariant mass region of the $\pi^0$, that stem
from the decay of the neutral pion. However, in the mass region
$M(\pi^0 \gamma)$ = 0.65--0.75 GeV there might be essential
``background'' with 3 (non-resonant) $\gamma$'s in the final state or
from four photon final states (e.g.\ the decay products from 2$\pi^0$,
$\eta \pi^0$ etc.) when one photon is not identified due to the finite
geometrical acceptance of the detector. In the following a method is
outlined which will help to identify events where the $3 \gamma$'s in
the final state are, in fact, produced via $\omega$-mesons.

It is possible to exploit, for example, kinematical conditions. When
an $\omega$-meson is produced not far above threshold it will have
comparatively small transverse momentum $P_t$ in the lab. system.  In
this case we can expect that the $\pi^0$ and $\gamma$ from the
$\omega$ decay will be strongly correlated in their transverse
momenta, while $\pi^0 \gamma $ events from the background will not
show such a correlation. In order to investigate this momentum
correlation in the case of $\omega$ decays inside the nucleus we
present in Fig. \ref{fig:om5} the distributions in the azimuthal angle
between two planes $\phi$ = $\phi_{\pi ^0}- \phi_{\gamma}$, where one
plane is formed by the initial proton momentum and the final $\pi ^0$
momentum (${\bf p}_0 \wedge {\bf p}_{\pi ^0}$) and the second one is
formed by the initial proton momentum and the $\gamma$ momentum (${\bf
  p}_0 \wedge {\bf p}_{\gamma}$). The solid histograms in Fig.
\ref{fig:om5} describe the distributions $dN/d\phi$ from $p + Pb$
(upper part) and $p + ^{12}C$ collisions (lower part) at $T_p = 2.4$
GeV (left) and 1.9 GeV (right) for 'inside' $\omega$ decays.  The
hatched histograms, which describe events with rescattered pions, are
essentially flat and do not show any correlation between the $\pi^0$
and $\gamma$. At the same time the solid histograms have distinctive
maxima at $\phi =$ 180$^o$ which correspond to correlated $\pi^0$ and
$\gamma$ events from the $\omega$ decay. The width of the distribution
in $dN/d\phi$ depends on the $\omega$-meson transverse momentum $P_t$.
Therefore it becomes more narrow with decreasing initial energy $T_p$;
very close to the threshold it will depend essentially on the Fermi
motion of the target nucleons and effects from $\omega$ rescattering.
These rescattering effects are smaller for $^{12}C$ than for $Pb$
(compare upper and lower solid histograms in Fig.  \ref{fig:om5}).
Nevertheless, the azimuthal correlation remains quite pronounced also
for $Pb$ even at $T_p$=2.4 GeV.

In Fig.  \ref{fig:om6} we present two dimensional plots for the
distribution in the azimuthal angle $\phi$ versus the $\omega$
transverse momentum $P_t$ (upper part) and in the invariant mass
$M(\pi^0, \gamma)$ versus $P_t$ (lower part) for $p+Pb$ at $T_p =$ 2.4
GeV. The l.h.s. describes events, where the rescattering of pions is
switched off, while in the r.h.s. only events with rescattered pions
are selected. By comparing the left and right distributions in Fig.
\ref{fig:om6} we conclude that the relative contribution of the
background from $\pi ^0$ rescattering can be essentially suppressed
choosing the cut $P_t \leq$0.2 GeV/c. This is demonstrated in more
detail in Fig.  \ref{fig:om8} (for $p+Pb$) and Fig. \ref{fig:om9} (for
$p+C$) at $T_p$=2.4 GeV, where we present the ratio

\begin{equation}
  R=\frac{N(M(\pi ^0,\gamma )=0.6-0.75~GeV)}{N(M(\pi ^0,\gamma ) \geq
    0.75~GeV)}
\end{equation}
calculated without cuts for the left histograms and with cuts on the
total and transverse momentum of the $\pi^0 \gamma $ system $P_{tot}
\leq $0.5 GeV/c, $P_t\leq $0.2 GeV/c and the azimuthal angle $\phi $ =
$\phi _{\pi ^0}-\phi _\gamma $= $180\pm 30^o$ for the right
histograms. Without cuts the ratio $R$ for $Pb$ ($C$) changes from
0.029 (0.008) -- when medium effects are absent (upper left
histograms) -- to 0.053 (0.020) when only collisional broadening is
taken into account (middle left histograms) and reaches 0.128 (0.059)
when both medium effects -collisional broadening and mass shift - are
included (lower left histograms). The calculations with cuts give the
ratio $R$=0.033
(0.020) without medium effects (upper right histograms), 0.075 (0.032) with $%
\delta \Gamma \neq 0$ but without mass shift (middle right
histograms) and 0.401 (0.183) with collisional broadening $\delta
\Gamma \neq 0$ and mass shift $\delta M\neq 0$ (lower right),
respectively. Therefore, using cuts on the total and transverse
momentum of the $\pi^0 \gamma $ system ( $P_{tot} \leq $0.5 GeV/c,
$P_t\leq $0.2 GeV/c) and the azimuthal angle
 $\phi $ = $\phi _{\pi ^0}-\phi _\gamma $= $%
180\pm 30^o$ it is possible to increase essentially the relative
contribution of medium effects and to suppress the background from
$\pi^0 $ rescattering or related uncorrelated sources. We point
out again, that the cut in the azimuthal angle $\phi$ is important
to suppress different sources of background which contain
uncorrelated $\pi^0 \gamma$ pairs.

The energy dependence of the $\omega$ production cross cross on
$Pb$ (upper figure) and C (lower figure) multiplied by the
branching BR($\pi^0 \gamma$) is shown in Fig. \ref{fig:om7}. The
dotted and dash-dotted curves correspond to $\omega$ decays
'outside' and 'inside' the nucleus, respectively, where the
'inside' component corresponds to events with a neutral pion (and
photon) in the final state. It thus does not include $\pi^0$
absorption or pion charge exchange reactions. The solid lines
describe the total $\omega$ yield in the final $\pi^0 \gamma$
channel, while the dashed lines are calculated for the events with
pion elastic rescattering from 'inside' $\omega$ decays. The total
cross section increases from 20 (3) $\mu$b at $T_p$ =1.7 to about
300 (40) $\mu$b at 2.6 GeV for a $Pb$ ($^{12}C$) target whereas
the ratio of the 'inside' to 'outside' component of the $\omega$
decay slightly decreases with bombarding energy due to kinematics.
Since this ratio  as well as the relative amount of $\pi^0$
elastic rescattering changes only smoothly with bombarding energy,
experiments at the highest energy of $T_p$ = 2.6 GeV are clearly
favored.

The target mass (A)-dependence of the $\omega$ production cross
section in the final $\pi^0 \gamma$ channel at 2.4 GeV is
presented in Fig. \ref{fig:om10}, where the notation of the curves
is the same as in Fig. 10. Here the ratio of 'inside'
(dash-dotted) to 'outside' (dotted) $\omega$ decays increases
substantially with mass number $A$, while the relative $\pi^0$
rescattering increases with $A$ only moderately. This effect
correlates with the average propagation time of the $\omega$ meson
in the nucleus.

It is important to note that the actual mass shift of the
$\omega$-meson at finite density is not known. In Eq.
(\ref{Brown}) we have used a coefficient $\alpha$ = 0.18 as
suggested by QCD sum rules \cite{3} for $\omega$-mesons at rest in
the nucleus. However, the $\omega$ self energy might well be a
function of momentum as suggested by the studies in Refs.
\cite{Wamb,Friman,Kondratrho} in case of the $\rho$ meson.
Therefore, the actual mass shift seen in $p- A$ reactions might be
different; however, when gating on different momentum intervals
for the $\omega$-mesons in the lab. this information might be
determined experimentally, too.

One of the most important source of three photon background is related
to the reactions $pp \to p \Delta^+ (p \gamma)\pi_0 (2 \gamma)$ and $p
N \to p \Delta^0 (n \gamma)\pi_0 (2 \gamma)$. Taking into account that
$\sigma (NN \to N \Delta \pi)$ can be about 1 mb and the branching
BR($\Delta \to N \gamma \simeq 5 \times 10^{-3}$ ) we find that
$\sigma _{BG} \simeq 5~ \mu$b per nucleon. At the same time the cross
section $\sigma_0 $ for the $\pi^0 \gamma$ yield from the $\omega$
decays inside the nucleus can be written as
\begin{equation}
\sigma_0 = \sigma (pN \to pN \omega)\cdot BR(\omega \to \pi^0
\gamma) \cdot P(\mbox{inside)},
\end{equation}
where $ \sigma (pN \to pN \omega) \simeq 200-300\mu$b (cf.
\cite{Nakayama}), $BR(\omega \to \pi^0 \gamma =0.085$ while the
probability $P_{Pb}(\mbox{inside)}$ for the $\omega$ to decay inside
the $Pb$ nucleus is $ \simeq 0.4$ (cf. Fig. 2). As a result we have
for a $Pb$ target $\sigma_0 \leq 7-10 \mu$b per nucleon. This implies
that the background roughly is of the same order than the signal from
$\omega$-production with subsequent decay in the nucleus. However, the
non-resonant background has broader angular and invariant 3$\gamma$-mass
distributions. Thus the analysis procedures outlined above can also be
applied for events outside the $\omega$ peak and in such a way ---
although not explicitly shown here -- it should be possible to
subtract the non-resonant events from ``real'' $\omega$-events with
decay in the nucleus.

\section{Summary}

In this work we have analyzed the possibility to detect in-medium
effects for the $\omega$-meson in the reaction $pA \to \omega
(\pi^0 \gamma) X$ by detecting 3 photons in the final channel
where 2 photons add up in invariant mass to the $\pi^0$ mass. As
already shown in Ref. \cite{Sibirtsev}, the rescattering of pions
will not spoil completely the in-medium effect related to the
$\omega$-mass shift. Furthermore, we have demonstrated here that
the in-medium $\omega$ decay to off-shell pions (and $\gamma$)
does not modify the reconstructed invariant mass distribution very
much relative to the dynamics of on-shell pions (cf. Fig. 1).
However, the uncertainty of missing one photon due to the limited
acceptance of the photon detector or possible background from
nonresonance production of $\pi^0 \gamma$ might create serious
problems for an unambiguous identification of the $\omega$ decay
products.

For the kinematical conditions of the spectrometer ANKE at the
proton accelerator COSY (J\"ulich) the $\omega$ meson can be
produced only close to threshold and therefore will have small
transverse momenta. This provides the possibility to separate a
signal from the $\omega$ decay from an uncorrelated $\pi^0 \gamma$
pair using back-to-back correlations in transverse momentum of the
$\pi^0$ and $\gamma$. Alternatively, one can use the distributions
in azimuthal angle between two planes $\phi$ = $%
\phi_{\pi ^0}- \phi_{\gamma}$, where one plane is formed by the
initial proton momentum and the $\pi ^0$ momentum (${\bf p}_0
\wedge {\bf p}_{\pi ^0}$) and the second one is formed by the
initial proton momentum and the $\gamma$ momentum (${\bf p}_0
\wedge {\bf p}_{\gamma})$. We have demonstrated that the
distribution in this azimuthal angle provides quite selective
criteria for a separation of the in-medium effect.

We also have investigated, how different kinematical cuts can
increase the signal-to-background ratio. The in-medium
modifications are found to be most pronounced for the cuts on the
total and transverse momentum of the $\pi^0 \gamma $ system
$P_{tot} \leq $0.5 GeV/c, $P_t\leq $0.2 GeV/c and in the azimuthal
angle $\phi $ = $\phi _{\pi ^0}-\phi _\gamma $= $%
180\pm 30^o$. For example, the ratio
$R =\frac{N(M(\pi ^0,\gamma )=0.6-0.75~GeV)}{N(M(\pi ^0,\gamma ) \geq 0.75~GeV)%
}$ calculated  for $Pb$ ($C$) at $T_p$ = 2.4 GeV without cuts
changes from 0.029 (0.008) -- when medium effects are absent -- to
0.128 (0.059) when both medium effects -collisional broadening and
mass shift - are included. The calculations with cuts give the
ratio $R$=0.033 (0.020) without medium effects and 0.401 (0.183)
with collisional broadening $\Delta \Gamma \neq 0$ and mass shift
$\Delta M\neq 0$ included. Therefore, the study of in-medium
effects for $\omega$-mesons via the $\pi^0 \gamma$ decay mode in
$pA$ collisions in the near threshold region appears very
promising.

\vspace{1cm} The authors acknowledge many useful discussions with
K. Boreskov, A. Dolgolenko  and A. Sibirtsev throughout this
study.

\newpage

\begin{figure}[t]
\phantom{a}\vspace*{-2.5cm}
\centerline{\psfig{figure=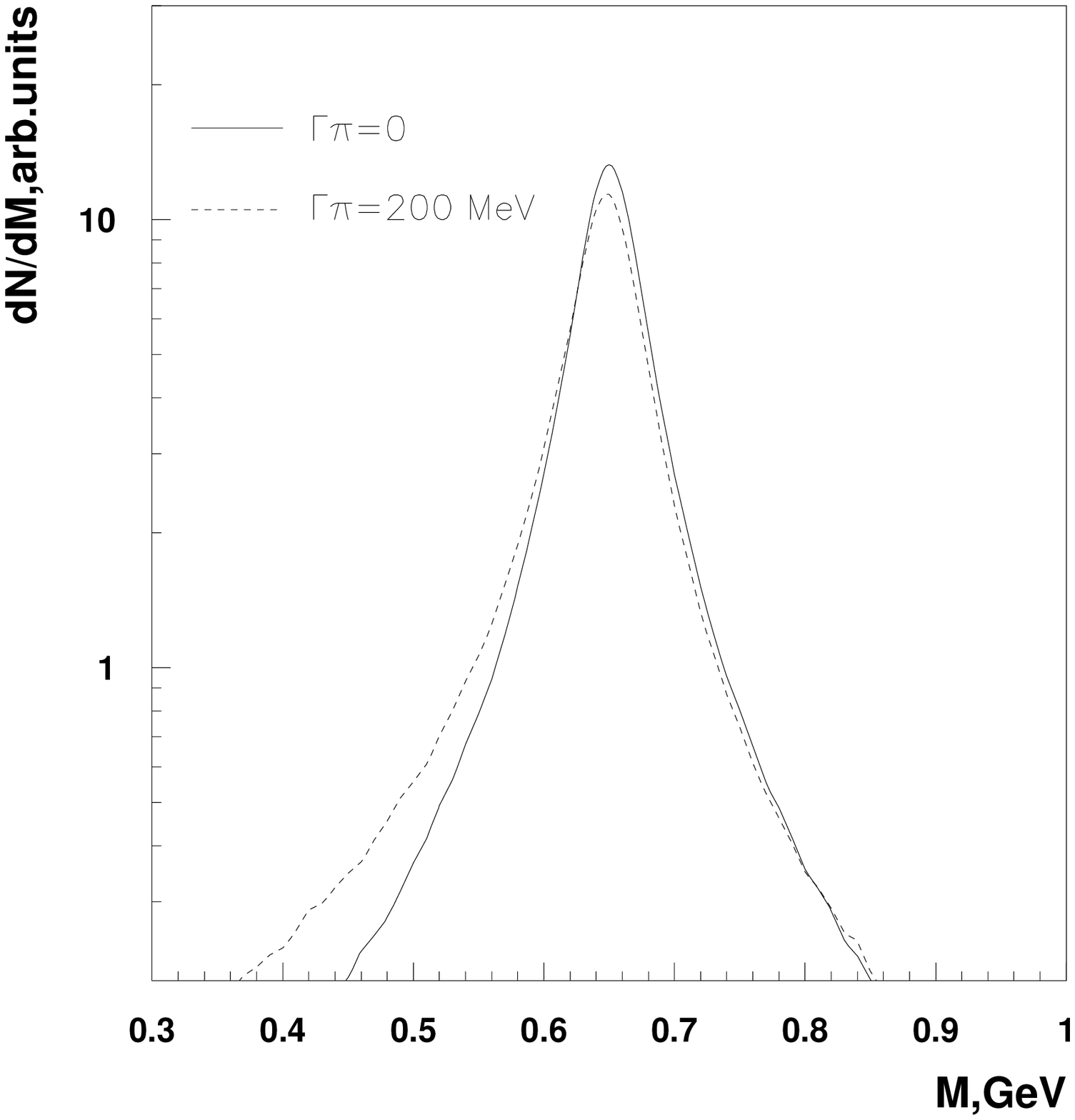,width=16cm}}
\vspace*{+1cm}
\caption{Invariant mass
distribution of in-medium $\omega$ decays to $\pi^0 \gamma$ as
reconstructed within on-shell pion dynamics (solid line) in
comparison to off-shell pion dynamics (dashed line) for a pion
spectral function with a collisional width of 200 MeV.}
\label{fig:om0}
\end{figure}

\begin{figure}[t]
\phantom{a}\vspace*{-2.5cm}
\centerline{\psfig{figure=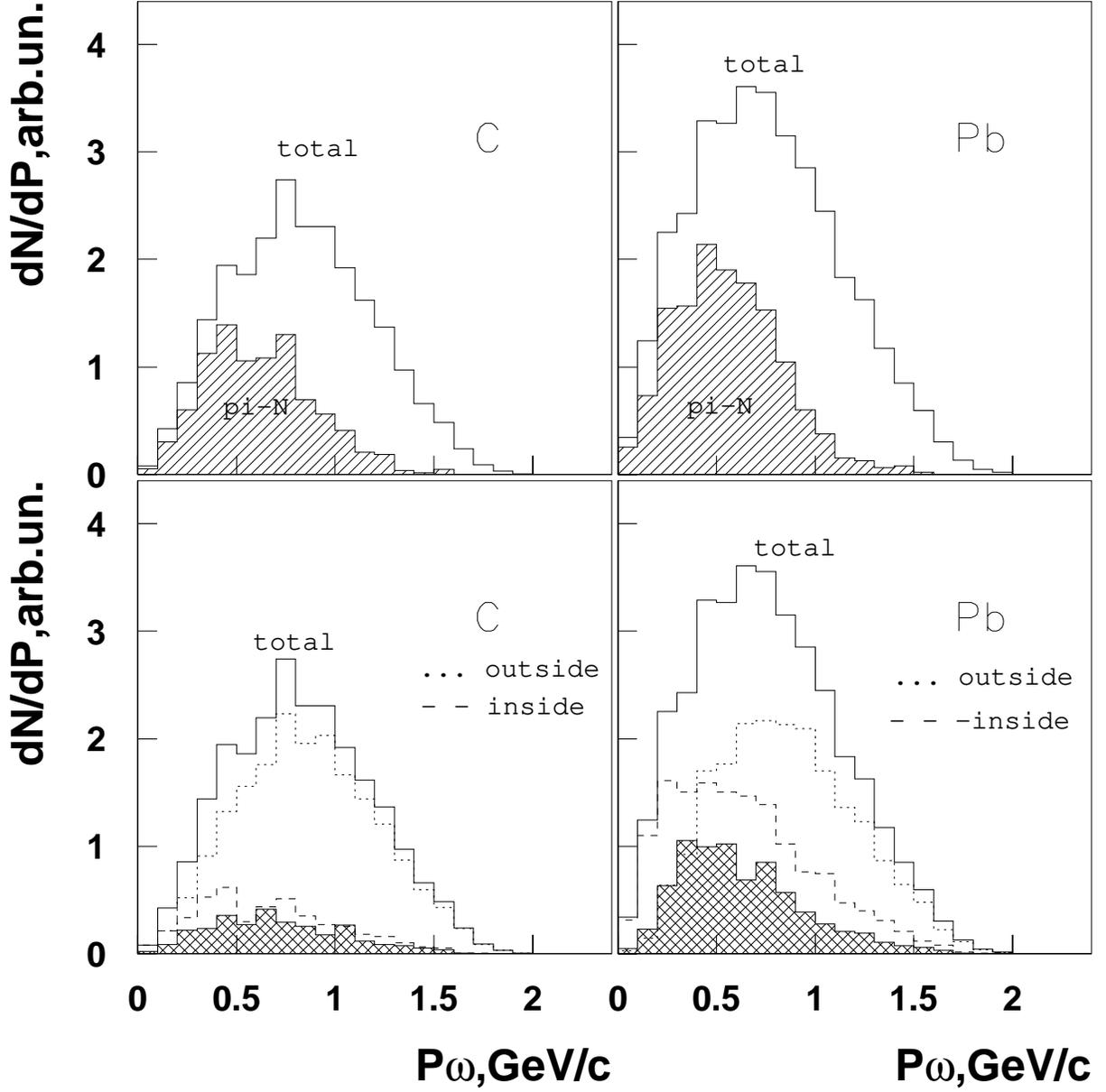,width=16cm}} \vspace*{1cm}
\caption{Momentum distributions of $\omega $ mesons in the
reactions $p\ +^{12}C$ (l.h.s.) and $p+^{208}Pb$ (r.h.s.) at
$T_p=$2.4 GeV (solid histograms). The contributions from the
two-step $\omega $ production mechanism is shown by the hatched
areas in the upper parts. The dotted and dashed histograms in the
lower parts correspond to $\omega $ decays 'outside' and 'inside'
the nucleus, respectively. The hatched areas in the lower figures
describe the contribution from the 'inside' component with
surviving pions, i.e. for pions from the $\omega$ decay that
propagate to the vacuum, however, may rescatter elastically.}
\label{fig:om1}
\end{figure}

\begin{figure}[t]
\phantom{a}\vspace*{-2.5cm}
\centerline{\psfig{figure=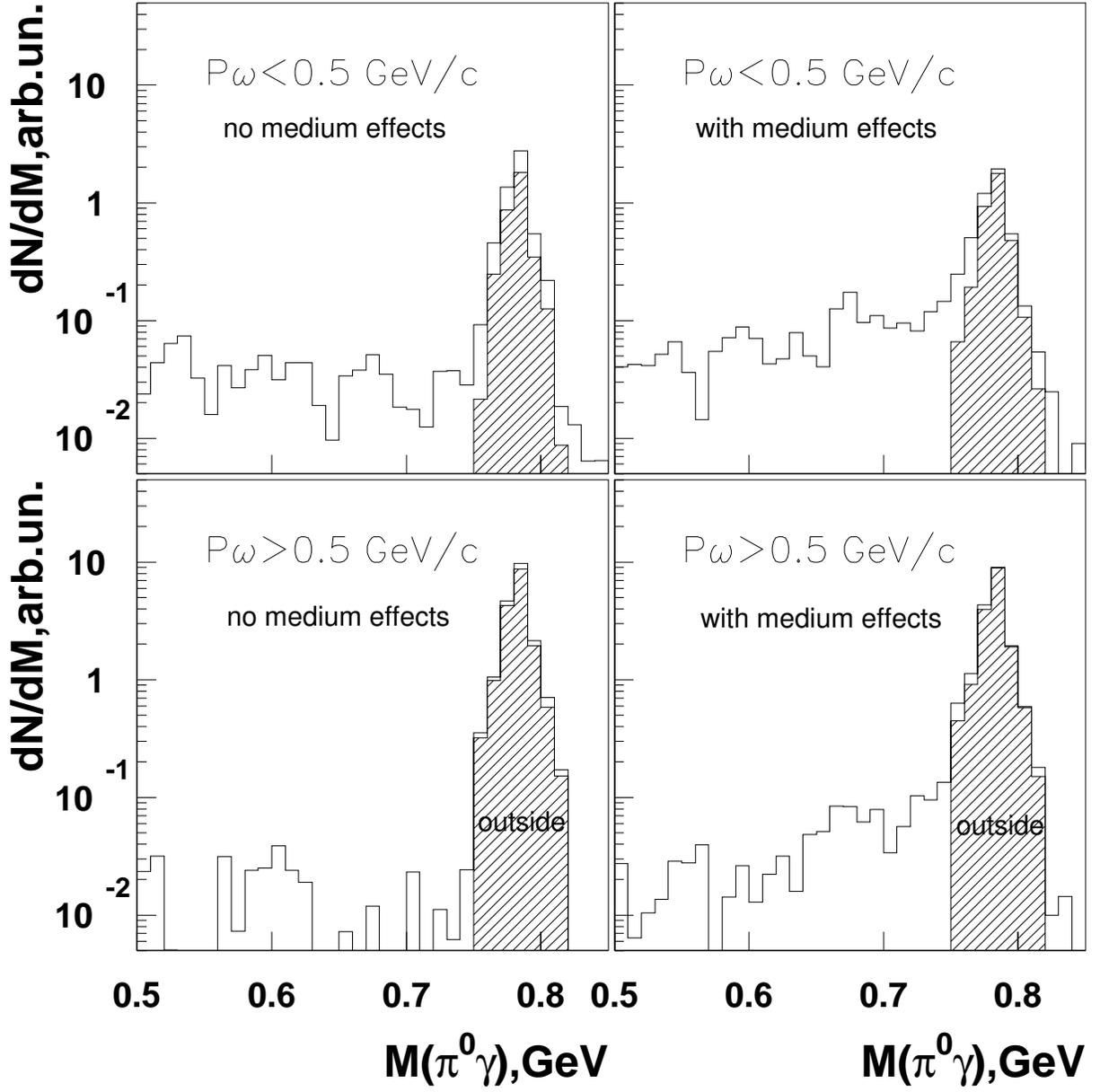,width=16cm}} \vspace*{+1cm}
\caption{The $\pi ^0\gamma $ invariant mass spectra for $p+Pb$
collisions at $T_p$ = 2.4 GeV. The upper and lower histograms are
calculated for  'slow' ( $p_\omega \leq $0.5 GeV/c) and 'fast' (
$p_\omega \geq $ 0.5 GeV/c) $\omega $-mesons, respectively. The
right (left) figures show the distributions with (without) medium
effects. The hatched areas correspond to $\omega $ decays outside
the nucleus.} \label{fig:om2}
\end{figure}

\begin{figure}[t]
\phantom{a}\vspace*{-2.5cm}
\centerline{\psfig{figure=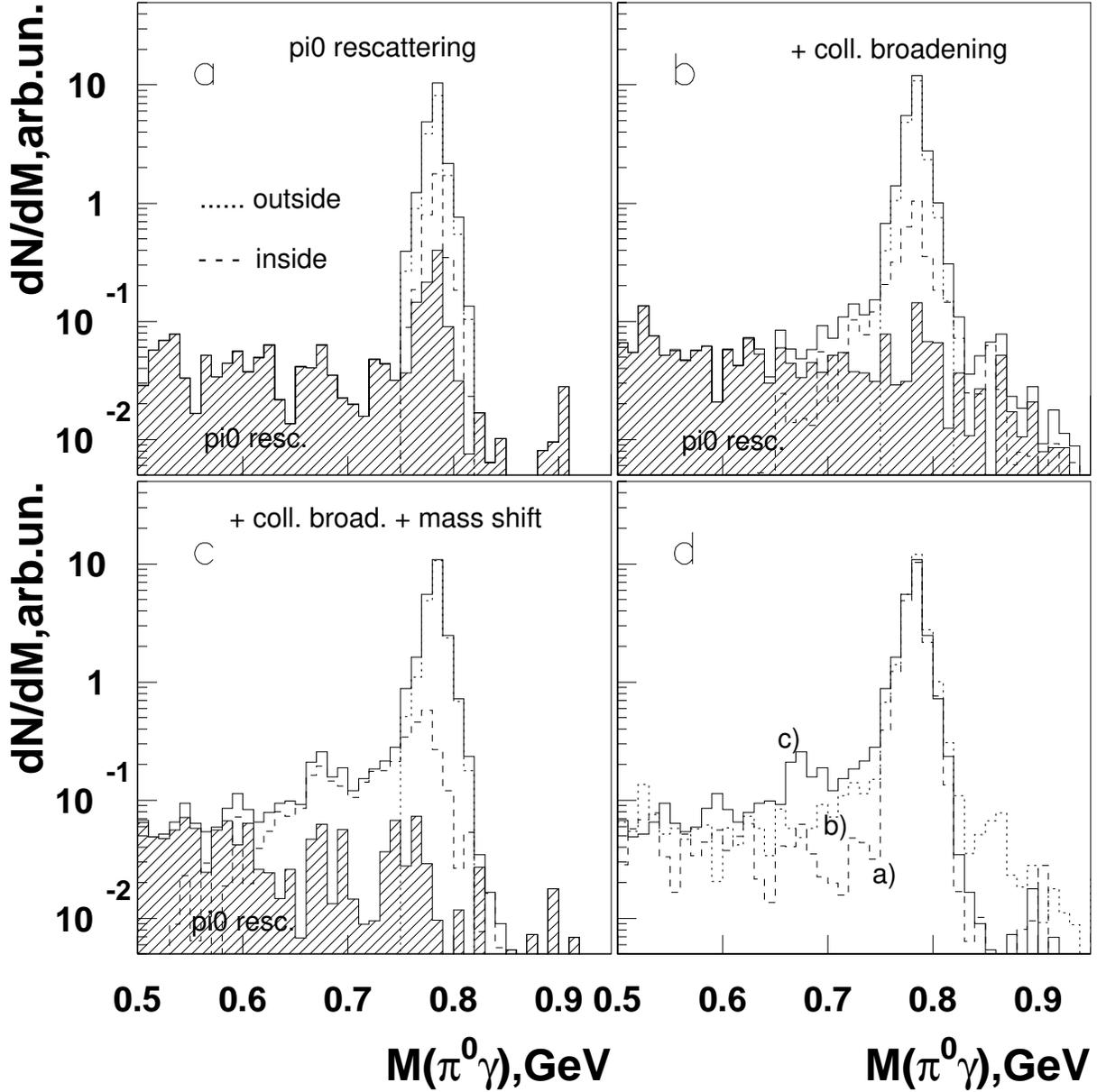,width=16cm}}
\vspace*{+1cm}
\caption{The
$\pi^0\gamma $ invariant mass distributions in $p+Pb$ collisions
at $T_p$ = 2.4 GeV for different scenarios: a) no $\omega$
in-medium effects are taken into account and all the distorsion of
the $M(\pi^0 \gamma)$ spectrum from the $\omega$ decay is caused
by $\pi^0$ rescattering; b) the collisional broadening $\delta
\Gamma$ is included for the $\omega$ spectral function but no mass
shift; c) both in-medium effects $\delta \Gamma$ and $ \delta M$
are taken into account; d) the dashed, dotted and solid curves
correspond to the solid curves in a), b) and c), respectively. The
hatched areas in a)-c) describe events with rescattered pions. The
dotted histograms in a)-c) show events with $\omega$'s decaying
outside, while the dashed histograms correspond to $\omega$'s
decaying inside the nucleus with their rescattering switched off.}
\label{fig:om3}
\end{figure}

\begin{figure}[t]
\phantom{a}\vspace*{-2.5cm}
\centerline{\psfig{figure=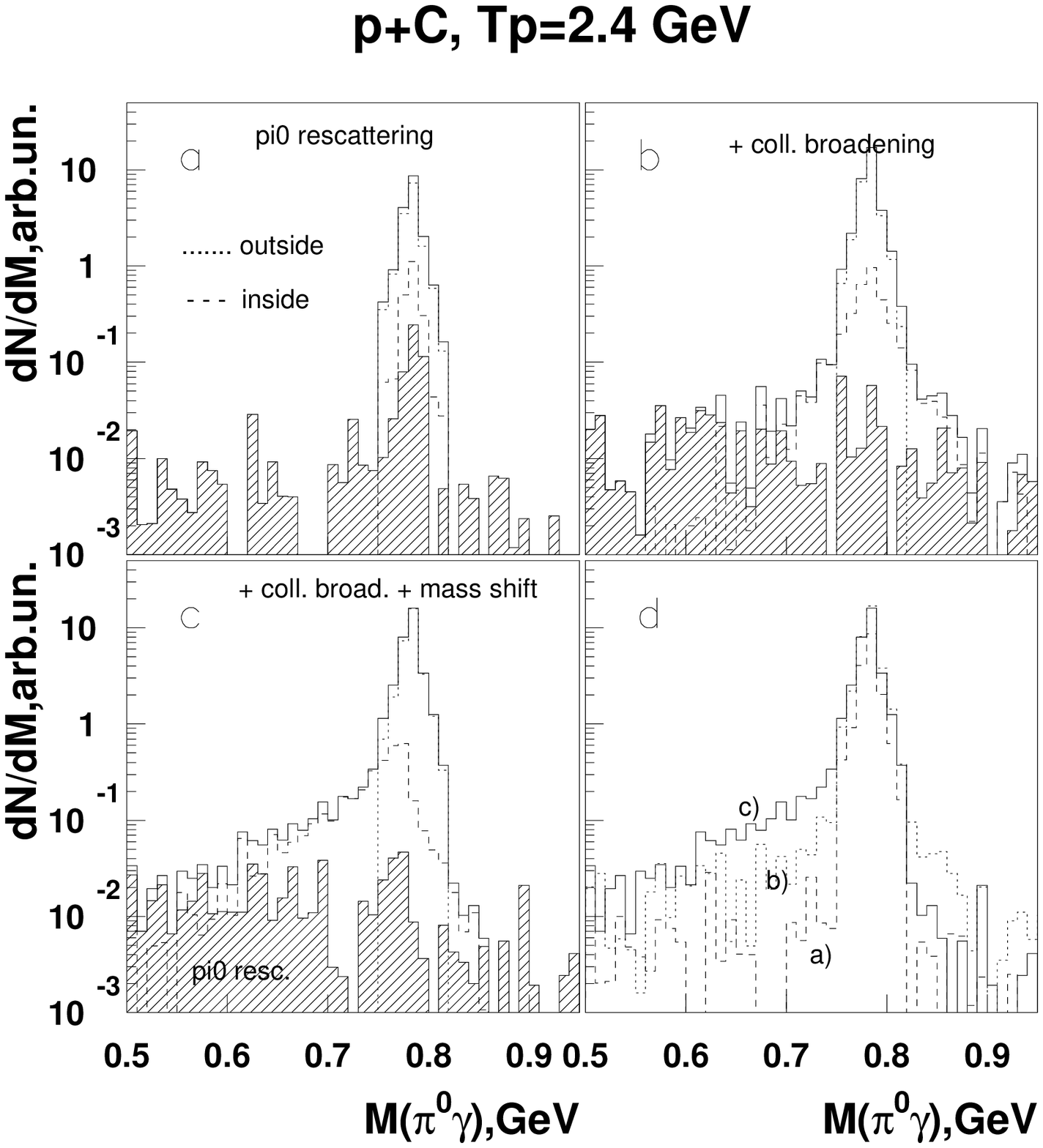,width=16cm}} \vspace*{+1cm}
\caption{The $\pi^0 \gamma $ invariant mass distributions in $p+C$
collisions at $T_p$ = 2.4 GeV, respectively. The meaning of the
histograms is the same as in Fig. 4.} \label{fig:om4}
\end{figure}

\begin{figure}[t]
\phantom{a}\vspace*{-2.5cm}
\centerline{\psfig{figure=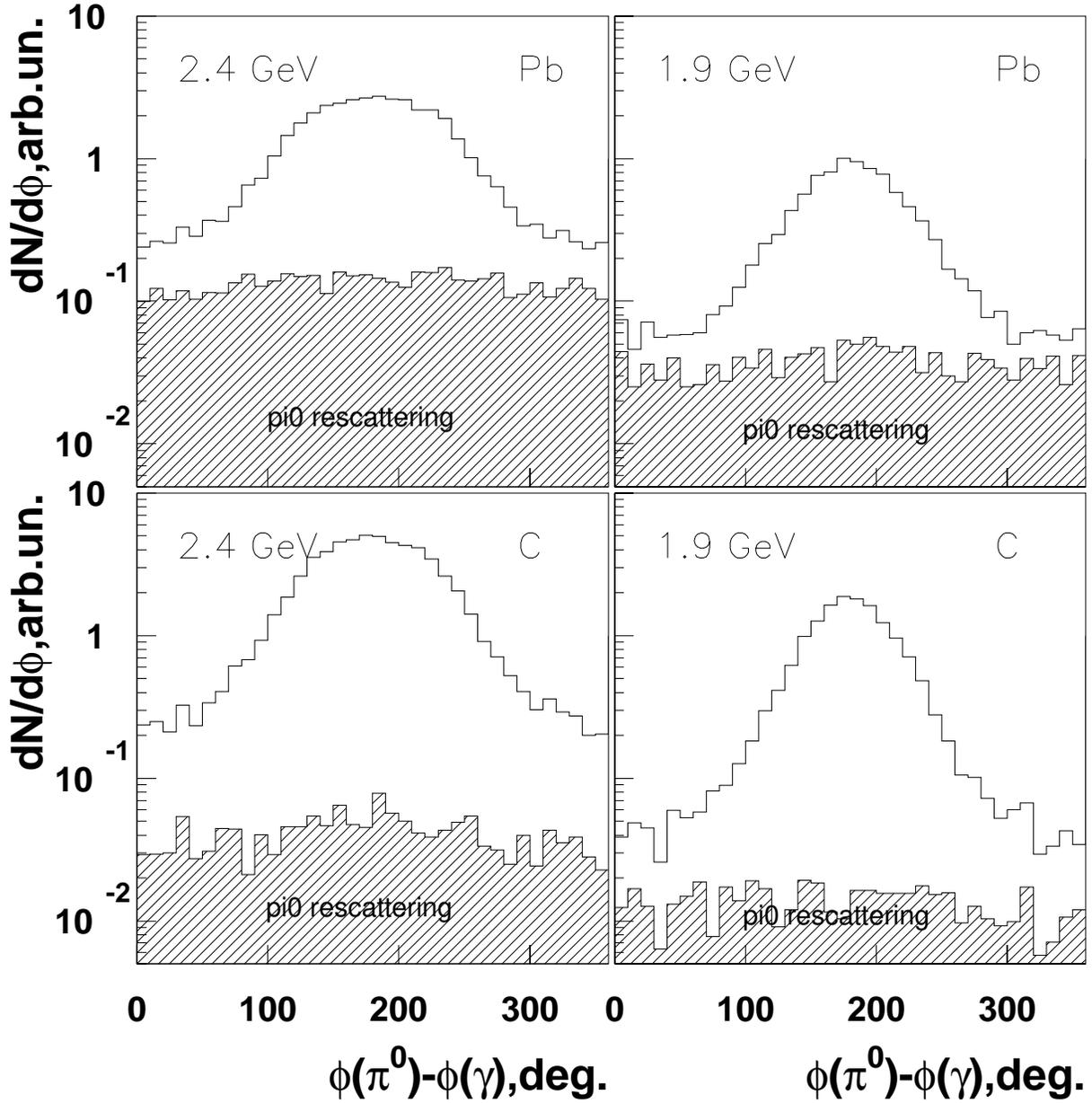,width=16cm}}
\vspace*{+1cm}
\caption{
The distributions in azimuthal angle between two planes $\phi$ = $%
\phi_{\pi ^0}- \phi_{\gamma}$, where one plane is formed by the
initial proton momentum and the final $\pi ^0$ momentum (${\bf
p}_0 \wedge {\bf p}_{\pi ^0}$) and the second one is formed by the
initial proton momentum and the $\gamma$ momentum (${\bf p}_0
\wedge {\bf p}_{\gamma}$). The solid histograms describe the
distributions $dN/d\phi$ from $p+Pb$ (upper part) and $p+C$
collisions (lower part) at $T_p = 2.4$ GeV (left) and 1.9 GeV
(right) for $\omega$'s decaying inside. The hatched histograms
show the events with rescattered pions. The solid histograms have
distinctive maxima at $\phi =$ 180$^o$ which correspond to
correlated $\pi^0$ and $\gamma$ pairs from the $\omega$ decay. }
\label{fig:om5}
\end{figure}

\begin{figure}[t]
\phantom{a}\vspace*{-2.5cm}
\centerline{\psfig{figure=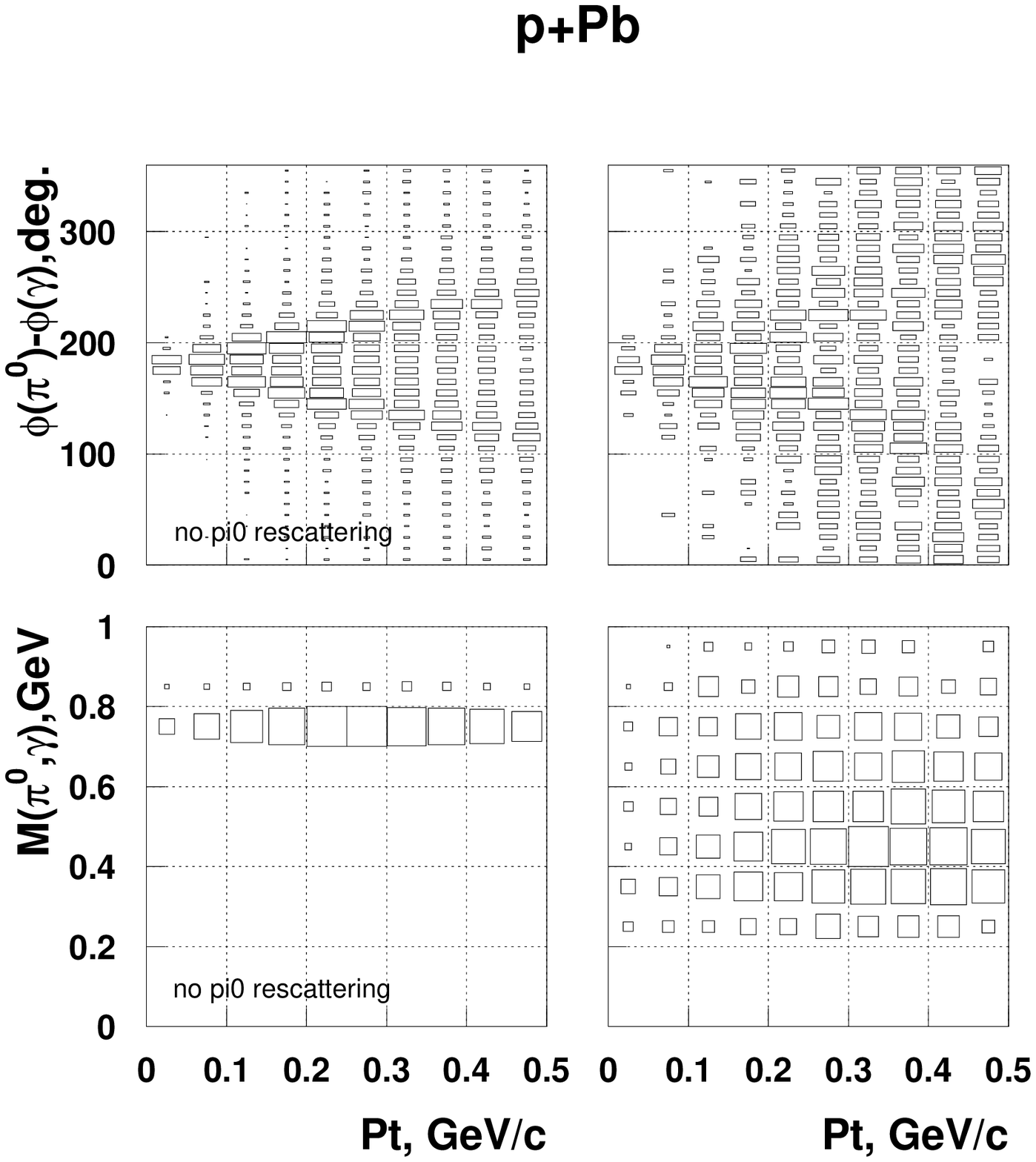,width=16cm}} \vspace*{+1cm}
\caption{ Two dimensional plots for the distribution in the
azimuthal angle $\phi$ versus the $\omega$ transverse momentum
$P_t$ (upper part) and the invariant mass $M(\pi^0, \gamma)$
versus $P_t$ (lower part) for $p+Pb$ at $T_p$ = 2.4 GeV. The left
plots describe the events, where the rescattering of pions is
switched off, while in the right plots only events with
rescattered pions are selected. } \label{fig:om6}
\end{figure}

\begin{figure}[t]
\phantom{a}\vspace*{-2.5cm}
\centerline{\psfig{figure=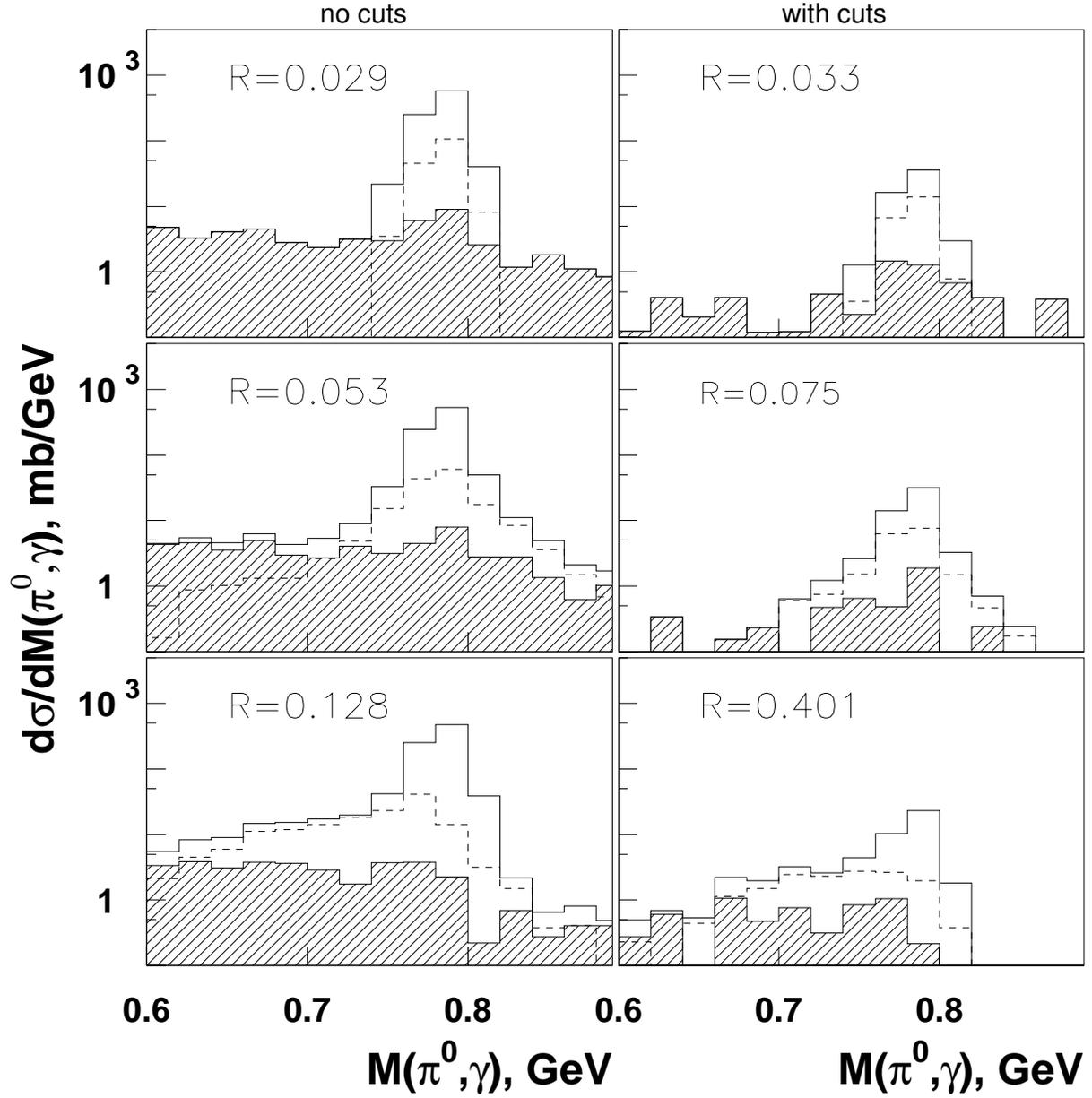,width=16cm}} \vspace*{+1cm}
\caption{ The ratio $R$ (20) for $p+Pb$ collisions calculated
without cuts (l.h.s.) and with cuts (r.h.s.) on the total and
transverse momentum of the $\pi^0 \gamma $ system  $P_{tot} \leq
$0.5 GeV/c, $P_t\leq $0.2 GeV/c and the azimuthal angle
 $\phi $ = $\phi _{\pi ^0}-\phi _\gamma $= $%
180\pm 30^o$.}
\label{fig:om7}
\end{figure}

\begin{figure}[t]
\phantom{a}\vspace*{-2.5cm}
\centerline{\psfig{figure=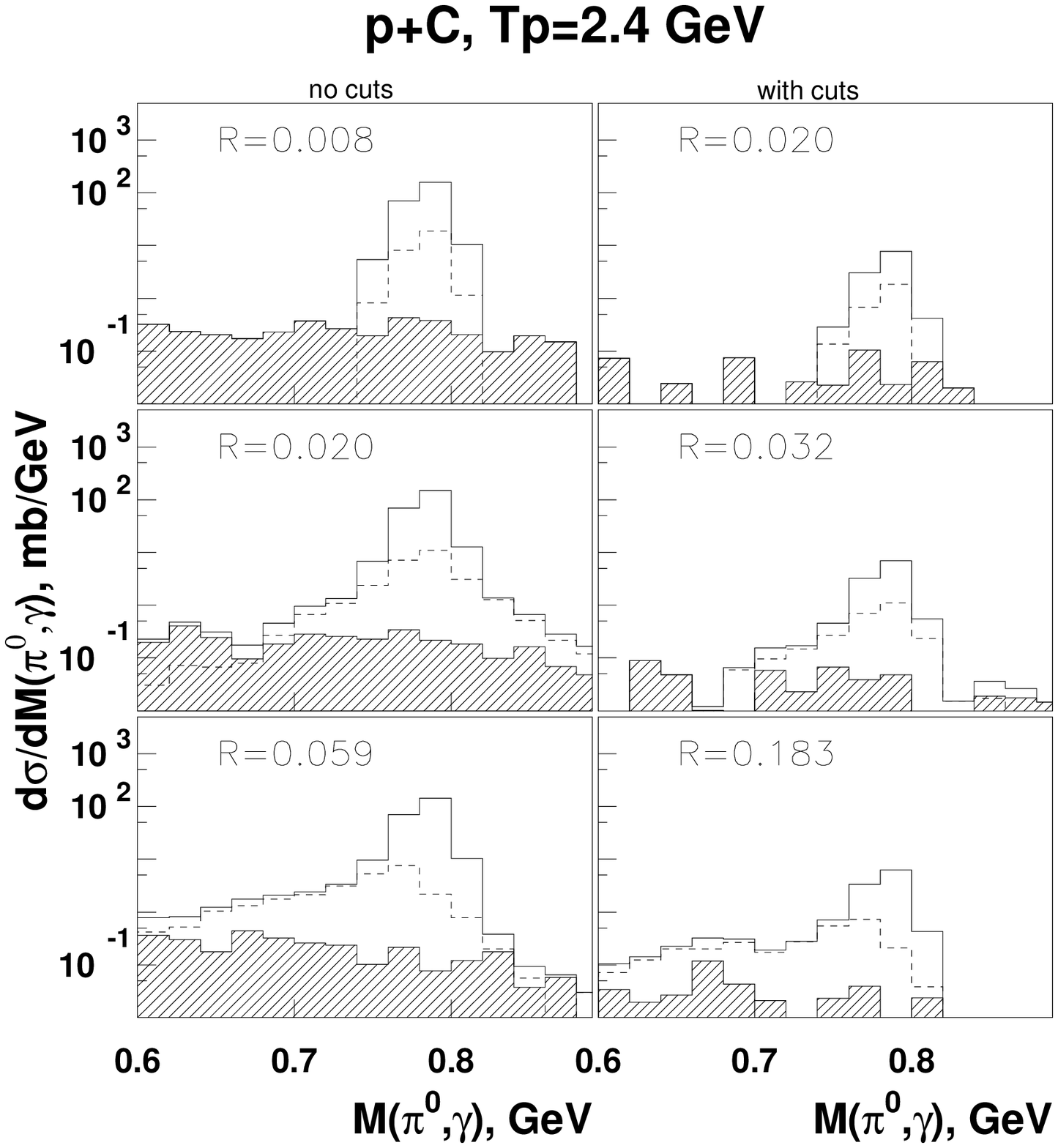,width=16cm}} \vspace*{+1cm}
\caption{ The same distributions as in Fig. 8  for $p+C$
collisions.} \label{fig:om8}
\end{figure}

\begin{figure}[t]
\phantom{a}\vspace*{-2.5cm}
\centerline{\psfig{figure=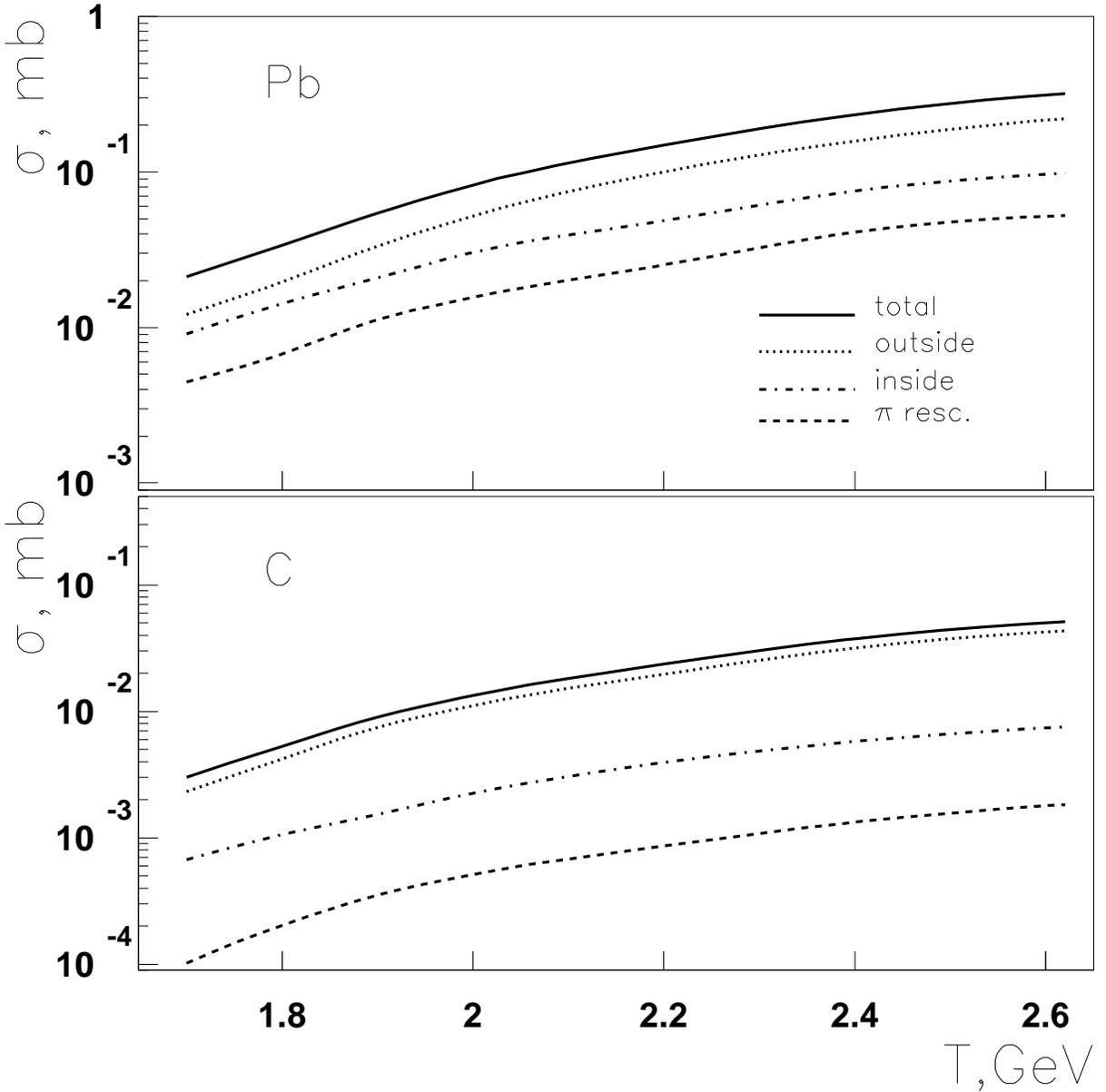,width=16cm}} \vspace*{+1cm}
\caption{ The energy dependence of the $\omega$ production cross
section on $Pb$ (upper part) and $^{12}C$ (lower part) multiplied
by the branching BR($\pi^0 \gamma$). The dotted and dash-dotted
lines correspond to $\omega$ decays 'outside' and 'inside' the
nucleus, respectively, where the 'inside' component corresponds to
events with a neutral pion (and photon) in the final state. It
thus does not include $\pi^0$ absorption or pion charge exchange
reactions. The solid curve describes the total $\omega$ yield in
the final $\pi^0 \gamma$ channel, while the dashed line reflects
events with pion elastic rescattering from the 'inside' component
of the $\omega$ decay.} \label{fig:om9}
\end{figure}

\begin{figure}[t]
\phantom{a}\vspace*{-2.5cm}
\centerline{\psfig{figure=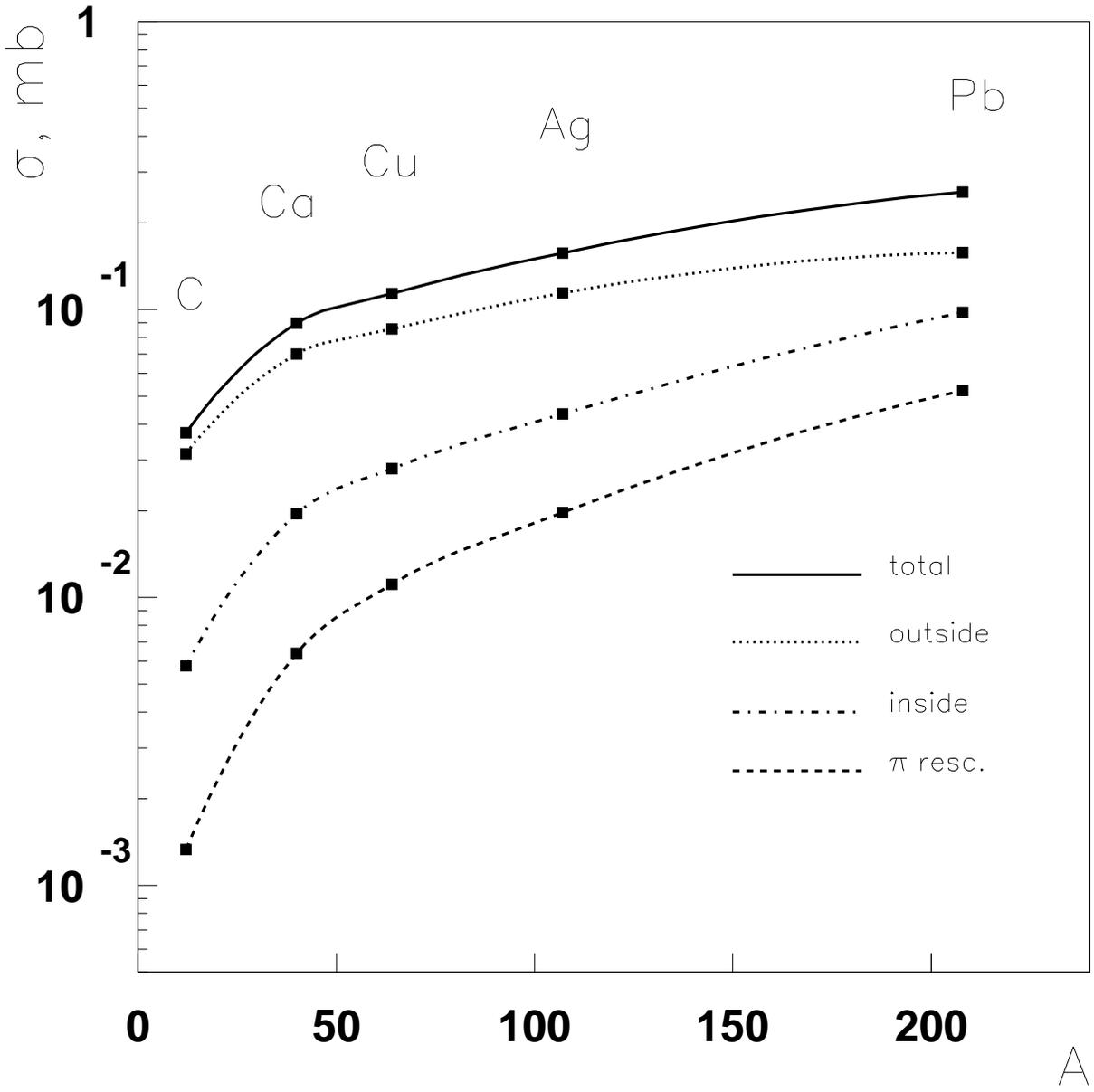,width=16cm}} \vspace*{+1cm}
\caption{ The target $A$-dependence of the $\omega$ production
cross section in the $\pi^0 \gamma$ channel at $T_p$ = 2.4 GeV.
The notation of the lines is the same as in Fig. 10.}
\label{fig:om10}
\end{figure}

\end{document}